\documentclass[aps,prl,twocolumn,
superscriptaddress,
showpacs,preprintnumbers,amsmath,amssymb,nofootinbib,longbibliography]{revtex4-1}
\usepackage[
colorlinks=true, 
pdfstartview=FitV, 
linkcolor=blue, 
citecolor=magenta, 
urlcolor=blue, 
bookmarks=true,
bookmarksnumbered=true,
pdftitle={},
pdfauthor={Masaru Hongo (RIKEN)}
]{hyperref}
\usepackage[dvips]{graphicx}
\usepackage{bm,latexsym,amsmath,amssymb,amsfonts,mathrsfs,textcomp}
\usepackage{color}
\input{colordvi.tex}

\newcommand\Lcal{\mathcal{L}}

\newcommand\eff{\mathrm{eff}}

\newcommand{\para}{\parallel}

\newcommand{\tr}{\mathop{\mathrm{tr}}}

\newcommand{\bnab}{\bm{\nabla}}
\newcommand{\bk}{\bm{k}}
\newcommand{\bx}{\bm{x}}

\newcommand{\bpi}{\bm{\pi}}

\newcommand{\bvarphi}{\bm{\varphi}}

\newcommand{\hrho}{\hat{\rho}}
\newcommand{\hU}{\hat{U}}

\newcommand{\average}[1]{\langle#1\rangle}
\newcommand{\baverage}[1]{\Big\langle#1\Big\rangle}
\newcommand{\ovl}[1]{\overline{#1}}

\newcommand{\diff}{\mathrm{d}}

\newcommand{\im}{\mathrm{i}}

\newcommand{\Rbb}{\mathbb{R}}

\newcommand{\with}{\quad\mathrm{with}\quad}
\newcommand{\Tr}{\mathrm{Tr}}
\newcommand{\open}{\mathrm{open}}

\newcommand\sect[1]{{\it #1.}---}

\begin{document}

\title{
Effective Lagrangian for
Nambu-Goldstone modes in nonequilibrium open systems
}

\author{Masaru Hongo}
\affiliation{Research and Education Center for Natural Sciences, 
Keio University, Yokohama 223-8521, Japan}
\affiliation{RIKEN iTHEMS, RIKEN, Wako 351-0198, Japan}

\author{Suro Kim}
\affiliation{Department of Physics, Kobe University, Kobe 657-8501, Japan}

\author{Toshifumi Noumi}
\affiliation{Department of Physics, Kobe University, Kobe 657-8501, Japan}
\affiliation{Department of Physics, University of Wisconsin-Madison, Madison, WI 53706, USA}

\author{Atsuhisa Ota}
\affiliation{Department of Applied Mathematics and Theoretical Physics,
University of Cambridge, Cambridge, CB3 0WA, UK}

\preprint{KOBE-COSMO-19-10}

\begin{abstract}
We develop the effective field theory of diffusive Nambu-Goldstone (NG) modes associated with spontaneous internal symmetry breaking 
taking place in nonequilibrium open systems.
The effective Lagrangian describing semi-classical dynamics of the NG modes is derived and matching conditions 
for low-energy coefficients are also investigated.
Due to new terms peculiar to open systems, the associated NG modes show diffusive gapless behaviors
in contrast to the propagating NG mode in closed systems.
We demonstrate two typical situations relevant to the condensed matter physics and high-energy physics, where diffusive type-A or type-B NG modes appear. 
\end{abstract}

\maketitle

\sect{Introduction}
Symmetry and its realization give fundamental descriptions of physical systems from condensed matter physics to high-energy physics.
Global symmetry of the system, if it exists, remains unbroken, or spontaneously broken in a given situation---e.\,g. in the ground state---and the resulting symmetry realization restricts a possible low-energy spectrum contained in the system: if spontaneous symmetry breaking (SSB) of continuous global symmetry takes place, it leads to the inevitable appearance of gapless excitations known as the Nambu-Goldstone (NG) modes according to the NG theorem~\cite{Nambu:1961tp,Goldstone:1961eq,Goldstone:1962es} with a low-dimensional exception protected by the Mermin-Wagner theorem~\cite{Mermin:1966fe,Hohenberg:1967zz,Coleman:1973ci}.

Although the original NG theorem is applicable to systems respecting the Lorentz symmetry in the ground state, the scope of its 
application has been just recently extended to several interesting directions.
One example is a generalization of the NG theorem to a nonrelativistic system.
In the absence of the Lorentz symmetry, there is generally a mismatch between the number of NG modes and the number of the broken symmetries, and the associated (so-called type-B) NG mode often shows a quadratic dispersion relation $\omega = a\bk^2 \,(a \in \mathbb{R})$~\cite{Nielsen:1975hm,Leutwyler:1993gf,Miransky:2001tw,Schafer:2001bq,Nambu:2004yia,Brauner:2010wm,Watanabe:2011ec,Hidaka:2012ym,Watanabe:2012hr,Watanabe:2014fva,Hayata:2014yga}.
Furthermore, the notion of the symmetry and its spontaneous breaking are also extended to nonequilibrium \textit{closed} systems \cite{Endlich:2012vt,Grozdanov:2013dba,Haehl:2015pja,Crossley:2015evo,Haehl:2015uoc,Jensen:2017kzi,Glorioso:2017fpd,Haehl:2018lcu,Jensen:2018hse} and \textit{open} systems~\cite{Minami:2018oxl,Sieberer:2015svu,Hongo:2018ant,Hayata:2018qgt}, where diffusive gapless modes appear.
To achieve these developments, one most powerful tool, the effective field theory (EFT)~\cite{Coleman:1969sm,Callan:1969sn,Weinberg:1978kz}, 
has been actively used.
One can apply EFT to show a nonrelativistic generalization of the NG theorem~\cite{Leutwyler:1993gf,Watanabe:2012hr,Watanabe:2014fva} and to describe nonequilibrium closed systems respecting conservation laws~\cite{Endlich:2012vt,Grozdanov:2013dba,Haehl:2015pja,Crossley:2015evo,Haehl:2015uoc,Jensen:2017kzi,Glorioso:2017fpd,Haehl:2018lcu,Jensen:2018hse}.
Dissipative effects in the low-energy spectrum are captured with the help of the Schwinger-Keldysh (real-time) formalism~\cite{Schwinger:1960qe,Keldysh:1964ud,Chou:1984es}, and the doubled symmetry structure inherent in it plays a central role.

In this Letter, taking one step further, we develop EFT for the NG modes resulting from spontaneous symmetry breaking in nonequilibrium \textit{open} systems (or non-Hermitian systems) where the conservation law is violated by considering e.\,g. the diffusive coupling between the system and environment~\cite{Minami:2018oxl,Sieberer:2015svu,Hongo:2018ant,Hayata:2018qgt}.
Generalizing the Callan-Coleman-Wess-Zumino's (CCWZ) coset construction~\cite{Coleman:1969sm,Callan:1969sn} to the Schwinger-Keldysh formalism, we lay out a solid basis to construct the general effective Lagrangian for the open system NG modes, and apply it to two typical situations where the type-A and type-B NG modes appear.

\sect{Symmetry structure in open systems}
Real-time dynamics of quantum systems can be systematically described by the use of the Schwinger-Keldysh formalism~\cite{Schwinger:1960qe,Keldysh:1964ud,Chou:1984es}.
One most basic quantity is the closed-time-path generating functional (CTPGF) defined by 
\begin{equation}
 \begin{split}
  Z [j_1,j_2] 
  &\equiv \Tr
  \left( \hrho_0 
   \hU^\dag_{j_2} (\infty,-\infty) \hU_{j_1} (\infty,-\infty) \right)
  \\
  &= \int \mathcal{D} \varphi_1 \mathcal{D} \varphi_2
  e^{\im (S[\varphi_1;j_1] - S[\varphi_2;j_2])}
  \rho_0 [\varphi],
 \end{split}
  \label{eq:CTPGF1}
\end{equation}
where $\hU_j(t_2,t_1)$ denotes the time-evolution operator from $t_1$ to $t_2$ in the presence of the external field $j(t)$, $\hrho_0$ the initial density operator at $t= - \infty$.
In the second line, we used the path-integral expression for a system composed of $\varphi$ with its action $S[\varphi;j]$.
One crucial point for the Schwinger-Keldysh formalism is the doubled degrees of freedom $\varphi \to \{\varphi_1,\varphi_2 \}$ on the CTP.
As a result, if the system originally enjoys $G$-symmetry, it is also doubled; the phase weight $e^{\im (S[\varphi_1;j_1] - S[\varphi_2;j_2] )}$ is invariant under $(G_1 \times G_2)$-transformation acting on $\varphi_1$ and $\varphi_2$, 
respectively%
\footnote{
However, note that non-diagonal part of $(G_1 \times G_2)$ is explicitly broken due to the existence of the boundary, e.\,g. $\rho_0 [\varphi]$.
This broken symmetry is recently shown to be nonlinearly realized in the effective field theory for conserved systems such as dissipative hydrodynamics~(See e.\,g. \cite{Crossley:2015evo,Glorioso:2017fpd}). 
}.

The charges attached to that symmetry can be diffused if the system is put under the influence of environments, and, as a result, 
the above symmetry structure is modified.
For example, let us consider the total system composed of two kinds of dynamical variables $\varphi = \{\phi,\,\sigma\}$ with system variables $\phi$ and environment variables $\sigma$. 
After integrating out the environment variables, we obtain 
\begin{equation}
 Z [j_1,j_2] 
  = \int \mathcal{D} \phi_1 \mathcal{D} \phi_2 
  e^{\im S_{\open}[\phi_1,\phi_2;j_1,j_2]}.
\end{equation}
Then, the open system action $S_{\open}[\phi_1,\phi_2;j_1,j_2]$ is, in general, not invariant under the nondiagonal part of $(G_1\times G_2)$-transformation, which is a manifestation of violating conservation laws.
Even in that cases, the diagonal subgroup of doubled symmetry, which we call $G_A$-symmetry, still survives, and we can consider its spontaneous breaking in open quantum systems~\cite{Minami:2018oxl,Sieberer:2015svu,Hongo:2018ant,Hayata:2018qgt}.

Suppose that the spontaneous $G_A$-symmetry breaking of the quantum open system takes place. 
Integrating out gapped degrees of freedom, we would like to develop the low-energy EFT for the associated NG fields:
\begin{equation}
 Z [j_1,j_2] 
  = \int \mathcal{D} \pi_R \mathcal{D} \pi_A
  e^{\im S_{\eff} [\pi_R,\pi_A;j_1,j_2]}.
\end{equation}
Here we introduced the effective action $S_{\eff} = \int \diff^d x \Lcal_{\eff}$ for $\{\pi_R,\,\pi_A\}$---a combination of doubled NG fields in the so-called Keldysh basis whose properties will be elucidate in the next section.
The vital point here is that we need to pay attention to several basic restrictions to the CTPGF (e.\,g.\,the probablity conservation), which can be manifestly respected by demanding the following conditions for the effective action 
(See e.\,g. Refs.~\cite{Crossley:2015evo,Glorioso:2017fpd} in detail)
\begin{subequations}
 \begin{align}
  &S_{\eff} [\pi_R, \pi_A = 0] 
  = 0,
  \label{eq:SK1}
  \\
  &S_{\eff} [\pi_R,\pi_A]^*
  = - S_{\eff} [\pi_R,-\pi_A] ,
  \label{eq:SK2}
  \\
  &\mathrm{Im}\, S_{\eff} [\pi_R,\pi_A] 
  \geq 0,
  \label{eq:SK3}
 \end{align}
\end{subequations}
where we switched off the external field.
In the following, we will construct the effective action with vanishing external field $j$ based on the coset construction.

\sect{Coset and Maurer-Cartan 1-form}
Let us first specify building blocks of the effective Lagrangian attached to spontaneous symmetry breaking of open systems $G_A \to H_A$ with unbroken symmetry $H_A$\footnote{We employ the standard assumption in the coset construction that the broken generators are closed under the action of broken symmetry generators.}.
Na\"ively speaking, it may be natural to identify the associated NG fields as a coordinate of the coset $G_A/H_A$.
Nevertheless, due to the basic structure of the Schwinger-Keldysh EFT, we need to include the doubled NG fields into the effective Lagrangian.
With the help of a set of NG fields $\bpi = (\pi_R,\pi_A)$, we then introduce doubled cosets $\xi_i(\bpi)\in G_i/H_i~(i=1,2)$, where the $G_i$-transformation acts as
\begin{equation}
 \xi_i(\bpi)  
  \xrightarrow{G_i} 
  g_i \xi_i(\bpi) h_i^{-1}(\bpi,g_i) 
  ~\mathrm{with}~
  g_i\in G_i, ~ h_i\in H_i.
  \label{eq:G12Transf}
\end{equation}
Since we are considering the situation where all non-diagonal parts of $G_1 \times G_2$ are explicitly broken in the open system, we only need to respect the $G_A$-symmetry.
By choosing a special representative $\xi_{1,2} (\bpi) \equiv \xi (\pi_R) e^{\pm \im \pi_A/2}$ with $\xi (\pi_R) \equiv e^{\im \pi_R}$, which gives our definition of $\pi_R$ and $\pi_A$, we find the following simple transformation rule under $(g,g) \in G_A$-transfomation: 
\begin{equation}
 \begin{split}
 \xi (\pi_R) 
  &\xrightarrow{G_A}
  g \xi (\pi_R) h^{-1} (\pi_R,g), \\
  \pi_A 
  &\xrightarrow{G_A} 
  h (\pi_R,g) \pi_A  h^{-1} ( \pi_R,g) .
 \end{split}
 \label{eq:GATransfpi}
\end{equation}
In other words, we introduced the nonlinearly transforming NG field $\pi_R$ and the associated linearly transforming adjoint field $\pi_A$ under $G_A$-transformation.
Note that $\pi_R$ and $\pi_A$ indeed parametrize the average and the difference parts of the doubled NG fields, which demand the basic constraints \eqref{eq:SK1}-\eqref{eq:SK3}.
Then, projections of the Maurer-Cartan 1-form $ \alpha (\pi_R) 
  \equiv \im^{-1} \xi^{-1} (\pi_R) \diff \xi (\pi_R)$,
transform in the usual manner:
\begin{equation}
 \begin{split}
  \alpha_\perp (\pi_R)
  &\xrightarrow{G_A}  
  h \alpha_\perp (\pi_R) h^{-1} ,
  \\
  \alpha_\para (\pi_R)
  &\xrightarrow{G_A}
  h \alpha_\perp (\pi_R) h^{-1} + \im^{-1} h \diff h^{-1},
 \end{split}
\end{equation}
where $\alpha_\perp \equiv \tr (X_a \alpha) X^a $ and $\alpha_\para \equiv \tr (X_\alpha \alpha) X^\alpha $.
Among the generic generators $X_I$ of $G_A$ satisfying
\begin{equation}
 \begin{split}
  &[X_I, X_J] = \im X_K f^K_{~IJ}, 
  \quad
  \tr (X_I X_J) \equiv g_{IJ},
 \end{split}
\end{equation}
we introduced $X_a$ and $X_\alpha$ as generators which belong to broken and unbroken parts, respectively%
\footnote{
We will use the notation $I,J,\cdots$ to label generic generators of $G_A$, and $a,b,\cdots (\alpha,\beta,\cdots)$ for broken and (unbroken) ones.
}.
In the following, assuming $g_{a\alpha} = 0$, we lower (raise) the broken and unbroken indices by using the block diagonal part of the Cartan (inverse) metric $g_{ab}$ and $g_{\alpha\beta}$ ($g^{ab}$ and $g^{\alpha\beta}$).

On the other hand, it seems unnecessary to examine the explicitly broken $(g,g^{-1}) \in G_R$-transformation property.
Nevertheless, as will be shown later, we need them to clarify matching conditions for low-energy coefficients.
Restricting ourselves to the semi-classical description of the NG fields, we obtain the following $G_R$-transformation rule~\cite{SM1}:
\begin{equation}
 \begin{cases}
  \pi_R \xrightarrow{G_R}
  \pi_R + O (\hbar^2), 
  \\
  \pi_A \xrightarrow{G_R}
  \pi_A + 2 [e^{-\im\pi_R} \theta e^{i\pi_R}]_\perp + O (\hbar^3), 
 \end{cases}
 \label{eq:GRTransfpi}
\end{equation}
where we introduced $A_\perp \equiv \tr (X_a A) X^a $ with a transformation parameter $\theta \equiv \theta^I X_I$.
Note that we count $\pi_A$ and $G_R$-transformation parameter as $O(\hbar)$ as is usual for the Schwinger-Keldysh EFT~\cite{Crossley:2015evo,Glorioso:2017fpd}.

\begin{figure}[htb]
 \centering
 \includegraphics[width=0.8\linewidth]{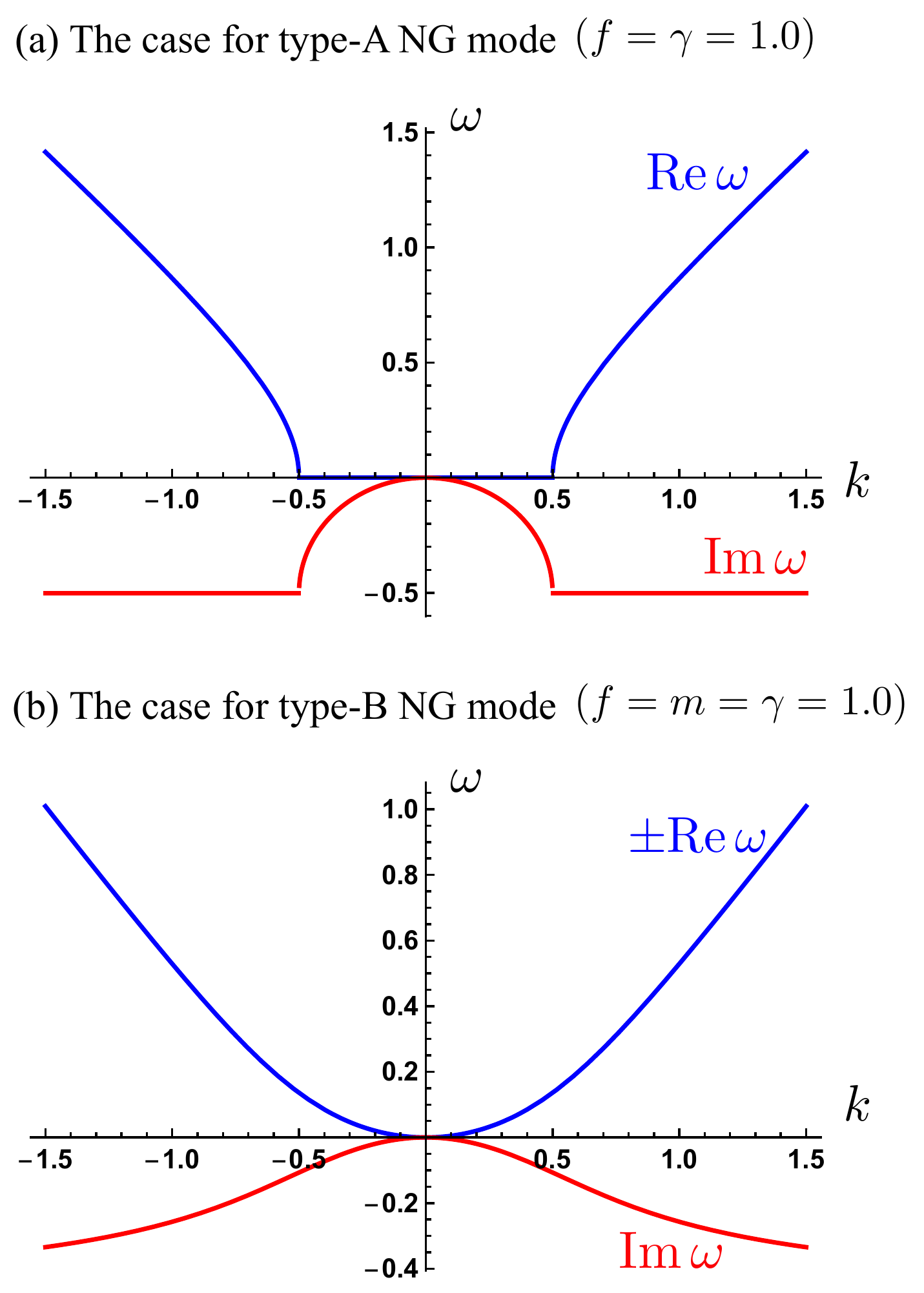}
 \caption{
 Dispersion relations for (a) type-A NG mode and (b) type-B NG modes with vanishing $\zeta$-terms.
 Blue (red) lines show the real (imaginary) part of the frequency.
} 
\label{fig:DR}
\end{figure}

\sect{Effective Lagrangian and diffusive NG modes}
Taking account of basic constraints of the Schwinger-Keldysh EFT~\eqref{eq:SK1}-\eqref{eq:SK3} and $G_A$-transformation properties developed in the previous section, we now construct the general $G_A$-invariant effective Lagrangian by the use of the $\alpha_{\perp,\para}$ and $\pi_A$ as basic building blocks.
Within the semi-classical level---including up to $O(\pi_A^2)$ terms, the leading-order $G_A$-invariant Schwinger-Keldysh effective Lagrangian is given as\footnote{For notational simplicity, we assume that the broken symmetry generators are irreducible under the unbroken symmetry transformation $H$. If not, the EFT parameters $\{f,m^\alpha,\gamma,\zeta_t^\alpha,\zeta_s^\alpha,A\}$ may have different values among irreducible sectors}
\begin{widetext}
 \begin{equation}
  \begin{split}
   \Lcal_{\eff} 
   = - F & \Bigg[
   \frac{1}{f^2} \tr ( \pi_A D_0 \alpha_{0\perp})
   + \im 
   \tr \big( m^{\alpha'} X_{\alpha'} [\pi_ A, \alpha_{0\perp}] \big)
   - \delta^{ij} \tr ( \pi_A D_i \alpha_{j\perp})
   \\
   & + \gamma \tr ( \pi_A \alpha_{0\perp}) 
   + \im 
   \tr \big( \zeta_t^{\alpha'} X_{\alpha'} [\pi_ A, D_0 \alpha_{0\perp}] \big)
   + \im \delta^{ij} \tr \big( \zeta_s^{\alpha'} X_{\alpha'} [\pi_ A, D_i \alpha_{j\perp}] \big)
   - \im \frac{A}{2F} \tr (\pi_A \pi_A)
   \Bigg] , 
   \label{eq:Leff1}
  \end{split}
 \end{equation}
\end{widetext}
where we defined a covariant derivative of $\alpha_{\perp}$ by
$ D_\mu \alpha_{\nu \perp} 
  \equiv \partial_\mu \alpha_{\nu \perp} 
  + \im [\alpha_{\mu\parallel}, \alpha_{\nu\perp}]$
which transforms covariantly:
$D_\mu \alpha_{\nu \perp} 
\xrightarrow{G_A} h D_\mu \alpha_{\nu \perp} h^{-1}$.
We here rescaled the NG fields as $\pi_{R,A} \to \pi_{R,A}/F$, and indices $\alpha'$ denotes the possible unbroken generators belonging to the center of $H_A$.
A set of real parameters $\{f,m^\alpha,\gamma,\zeta_t^\alpha,\zeta_s^\alpha,A\}$ gives low-energy coefficients, whose matching will be discussed shortly.
The condition~\eqref{eq:SK3} leads to $A \geq 0$, and when the charge is assumed to diffuse into the environment, we may also have $\gamma > 0$.
We note that it corresponds to (or defines) type-A or type-B NG mode whether terms proportional to $m^{\alpha'}$ vanish or not~\cite{Leutwyler:1993gf,Miransky:2001tw,Schafer:2001bq,Nambu:2004yia,Brauner:2010wm,Watanabe:2011ec,Hidaka:2012ym,Watanabe:2012hr,Watanabe:2014fva,Hayata:2014yga}.
Terms appearing in the second line of Eq.~\eqref{eq:Leff1} are not invariant under $G_R$-transformation. Hence, they are peculiar to the Schwinger-Keldysh EFT for nonequilibrium open systems, where $G_R$-symmetry is explicitly broken.
To see the quadratic part of the effective Lagrangian, expanding the Maurer-Cartan 1-form as 
$\alpha_{\mu\perp} = \partial_\mu \pi_R/F + O(\pi_R^2)$ and $\alpha_{\mu\parallel} = O (\pi_R^2)$, 
we obtain 
\begin{widetext}
 \begin{equation}
  \begin{split}
   \Lcal_{\eff} 
   &= - \rho_{ab} \pi_A^a \partial_0 \pi_R^b
   + g_{ab}^t \partial_0 \pi_A^a \partial_0 \pi_R^b
   - g_{ab}^s \bnab \pi_A^a \cdot \bnab \pi_R^b
   + \frac{\im}{2} g_{ab}^A \pi_A^a \pi_A^b
   \\
   &= \frac{\im}{2} 
   \begin{pmatrix}
    \pi_R^a & \pi_A^a
   \end{pmatrix}
  \begin{pmatrix}
   0 & \im ( g_{ab}^t \partial_0^2 - \rho_{ab} \partial_0 - g_{ab}^s \bnab^2) \\
   \im (g_{ab}^t \partial_0^2 + \rho_{ab} \partial_0 - g_{ab}^s \bnab^2) & 
   g_{ab}^A
  \end{pmatrix}
   \begin{pmatrix}
    \pi_R^b \\
    \pi_A^b
   \end{pmatrix},
  \end{split}
  \label{eq:EffectiveLag}
 \end{equation}
\end{widetext}
where we defined the following quantities
\begin{align}
 \rho_{ab} &\equiv 
 - m^{\alpha'} f_{\alpha' ab} + \gamma g_{ab} , 
 \quad
 g_{ab}^A \equiv A g_{ab},
 \\
 g_{ab}^t &\equiv
 f^{-2} g_{ab} - \zeta_t^{\alpha'} f_{\alpha' ab} 
 ,\quad 
 g_{ab}^s \equiv g_{ab} + \zeta_s^{\alpha'} f_{\alpha' ab}, 
\end{align}
and neglected surface terms resulting from the integration by parts.
We thus obtain the inverse retarded/advanced Green's functions for the NG modes as 
\begin{equation}
 (G_{R,A}^{ab})^{-1} (\omega, \bk) 
 = - g_{ab}^t \omega^2 \mp \im \rho_{ab} \omega + g_{ab}^s \bk^2.
 \label{eq:prop:RA}
\end{equation}
Solving $\det (G_R^{ab})^{-1} (\omega, \bk)  = 0 $ enables us to get the dispersion relation for the several types of NG modes depending on which parameters are present.
Fig.~\ref{fig:DR} shows two typical dispersion relations of the type-A and type-B NG modes with non-vanishing $\gamma$ and vanishing $\zeta$-terms.
We see the diffusive nature of the resulting NG modes 
due to the negative imaginary part $\mathrm{Im}\,\omega (\bk) < 0$.

Let us then specify the matching condition for low-energy coefficients in $\rho_{ab}$ and $g_{ab}$'s
based on the linearized effective Lagrangian~\eqref{eq:EffectiveLag}.
In contrast to SSB in the ground state, we generally have new couplings peculiar to nonequilibrium open systems:
$\rho_{ab}$ can have symmetric components whereas $g_{ab}^t$ and $g_{ab}^s$ can have anti-symmetric components.
To clarify the matching condition, noting that the leading infinitesimal transformation of the NG fields are given by $\pi^b_{R,A} \to \pi^b_{R,A} + \epsilon_{A,R}^a ( \delta_{A_a,R_a} \pi_{R,A}^b+ O(\epsilon,\pi))$ with
\begin{equation}
 \begin{cases}
  \delta_{A_a} \pi_R^b = F \delta_a^b,
  \\
  \delta_{A_a} \pi_A^b =0,  
 \end{cases}
 \mathrm{and} \quad
 \begin{cases}
  \delta_{R_a} \pi_R^b = 0, \\
  \delta_{R_a} \pi_A^b = F \delta^a_b,
 \end{cases}
 \label{eq:Transf}
\end{equation}
we turn our attention to the leading part of Noether currents attached to the $G_{A,R}$-transformations:
\begin{equation}
 \begin{split}
  J^\mu_{a,A} 
  &= \begin{pmatrix}
      F (g_{ba}^t \partial_0 \pi_A^b - \rho_{ba} \pi_A^b)
         \\
      - F g_{ba}^s \bnab \pi_A^b
    \end{pmatrix}
  + \cdots
  , \\
  J^\mu_{a,R} 
  &= \begin{pmatrix}
     F ( f^{-2} g_{ab} \partial_0 \pi_R^b -
     m^{\alpha'} f_{\alpha' ab} \pi_R^b)
      \\
      - F g_{ab} \bnab \pi_R^b
    \end{pmatrix}
  + \cdots ,
 \end{split}
 \label{eq:Currents}
\end{equation}
where the ellipsis stands for higher-order terms with respect to $\pi_{R,A}$.
Here $J^\mu_{a,R}$ is defined by the $G_R$-invariant part of the effective Lagrangian [corresponding to terms in the first line of Eq.~\eqref{eq:Leff1}], and hence, not conserved in open systems.
Using these, we can identify the matching conditions, which are sorted into the first category~\cite{SM1}\footnote{
It should be noticed that the currents $J_{A,R}^\mu$ are defined in the IR EFT~\eqref{eq:Leff1}. While $J_{A}^\mu$ is equivalent to the UV theory one up to equations of motion for gapped UV modes that are integrated out, it is nontrivial to identify the UV operator equivalent to the IR current $J_{R}^\mu$ essentially because the $G_R$ symmetry is explicitly broken. Also, the second condition in Eq.~\eqref{eq:matching1} involves a $G_R$ transformation, which is again an obstruction to the UV-IR matching. Therefore, further studies are required for establishing the UV-IR matching conditions, leaving it for future works.}
\begin{equation}
 \begin{split}
  &\average{\delta_{A_a} \pi_R^b} \big|_{\pi=0}
  = F \delta_a^b, \quad 
  \average{\delta_{R_a} J^0_{b,A}} \big|_{\pi=0}
  = - F^2 \rho_{ab}, \\
  &G_{J_{a,R}^0 J_{b,A}^0} (\omega,\bk=0) \big|_{\omega=0} 
  - \frac{F^2}{\omega} m^{\alpha'} f_{\alpha' ab}
  = \im \frac{F^2}{f^2} g_{ab},
 \end{split}
 \label{eq:matching1}
\end{equation}
and the second one
\begin{equation}
 \begin{split}
  &C^0_{ab} (\omega,\bk = 0) \big|_{\omega=0}
  = - \im F^2 \gamma g_{ab}, 
  \\
  &\partial_\omega C^0_{ab} (\omega,\bk = 0) \big|_{\omega=0}
  = - F^2  \zeta_t^{\alpha'} f_{\alpha' ab}, 
  \\
  &\partial_{k^i} C^i_{ab} (\omega=0,\bk ) \big|_{\bk=0}
  = - F^2  \zeta_s^{\alpha'} f_{\alpha' ab}, 
 \end{split}
 \label{eq:matching2}
\end{equation}
where we defined 
\begin{equation}
 \begin{split}
  & G_{J_{a,R}^\mu J_{b,A}^\nu} (\omega,\bk) 
  \equiv 
  \int \diff^d x e^{\im \omega t - \im \bk \cdot \bx}
  \average{J_{a,R}^\mu (x) J_{b,A}^\nu (0)},
  \\
  &C^\nu_{ab} (\omega,\bk)
  \equiv 
  \int \diff^d x e^{\im\omega t - \im \bk \cdot \bx}
  \average{\partial_\mu J_{a,R}^\mu (x) J_{b,A}^\nu (0)}.
 \end{split}
\end{equation}
Here the angle bracket (with $\pi = 0$) represents the path integral without dynamics of the NG fields. 
Note that $\gamma$ and $\zeta$-terms are matched by $C^\nu_{ab} (\omega,\bk)$ 
in the second category containing nonvanishing $\partial_\mu J_{a,R}^\mu (x)$ peculiar to open systems.
While Eqs.~\eqref{eq:matching1}-\eqref{eq:matching2} give matching conditions including $\zeta$-terms, we will consider two models with vanishing $\zeta$-terms in the following.

\sect{Examples}
As an example for a diffusive type-A NG mode, we consider a driven-dissipative BEC system with $U(1)$ symmetry \cite{Sieberer:2015svu} whose Lagrangian reads
\begin{align}
 \Lcal_{\open}
 &=
 \phi_A^\dag 
 \Big( \im \partial_0 + \frac{\bnab^2}{2m} + \mu + \im \kappa 
 - \frac{g - \im \gamma_0}{2} |\phi_R|^2 
 \Big) \phi_R
 \nonumber
 \\ 
 \quad  &
 + (\mbox{h.c.})
 + \frac{\im (A + 2 \gamma_0 |\phi_R|^2)}{2} |\phi_A|^2 
 + O(\phi_A^3),
\end{align}
where we introduced $\phi_R \equiv (\phi_1 + \phi_2)/2$ and $\phi_A \equiv \phi_1 - \phi_2$ for the doubled Bosonic Schr\"odinger field.
When $\kappa<0$ and $\gamma_0>0$, a driven-dissipative condensate $\ovl{\phi}_R = v e^{- \im \omega_0 t}$ could arise.
Note that this solution is regarded as an example of \textit{time crystal} in nonequilibrium open systems~\cite{Minami:2018oxl,Sieberer:2015svu,Hongo:2018ant,Hayata:2018qgt} since it spontaneously breaks a mixed part of time-translation and $U(1)$ symmetry~\cite{SM2}.
In this case, terms proportional to $\zeta^{\alpha'}_t,\,\zeta^{\alpha'}_s$, and $m^{\alpha'}$ all vanish, and the dispersion relation for the resulting NG modes is obtained as
\begin{equation}
 \omega = - \im \dfrac{g}{ m \gamma_0} \bk^2 + O(\bk^4). 
\end{equation}
We thus see that there exists a diffusive gapless mode, whose dispersion relation corresponds to small $|k| (<k_c)$ part of Fig.~\ref{fig:DR}\,(a) with $k_c = f\gamma/2$ (the exceptional point).
Note that the number of the gapless modes agrees with that of broken symmetries; ${\mathrm{rank} (g_{ab})} = \dim G_A/H_A$.

On the other hand, the dissipative $SU(2) \times U(1)$ model with a chemical potential defined by
\begin{align}
  \Lcal_{\open}
  &= \bvarphi_A^\dag 
  \big[ - (\partial_0 + \im \mu)^2 +  \bnab^2 - \gamma_0 \partial_0 
  - 2 \lambda \bvarphi_R^\dag \bvarphi_R
  \big] \bvarphi_R
 \nonumber
 \\
 &\hspace{50pt}
 + (\mathrm{h.c.}) + \im A \bvarphi_A^\dag \bvarphi_A,
\end{align}
gives an example for the type-B NG mode~\cite{Minami:2018oxl}.
This model can be regarded as an effective model describing the Kaon condensation in the dense QCD matter~\cite{Miransky:2001tw,Schafer:2001bq} coupled to the environment. 
Here $\bvarphi_{R,A}$ denote doubled two-component complex scalar fields.
One can find a stationary solution for $\bvarphi_R$'s equation of motion parametrized by e.\,g. $\ovl{\bvarphi}_R = (0,v)$ with $v = \mu/\sqrt{2\lambda}$, which spontaneously breaks $G_A =SU(2) \times U(1)$ symmetry down to $H_A =U(1)$.
Due to a nonvanishing anti-symmetric part of $\rho_{ab}$, this system contains the type-B NG mode and its gapped partner, whose dispersion relation is shown to be
\begin{equation}
 \begin{cases}
  \omega = \dfrac{\pm m - \im \gamma}{m^2 + \gamma^2} \bk^2
  + O (\bk^4), 
  \\
  \omega = f^2 ( \pm m - \im \gamma) 
  + \dfrac{ \pm m + \im \gamma}{m^2 + \gamma^2} \bk^2
  +O(\bk^4).
 \end{cases}
\end{equation}
This gapless diffusive-propagating behavior of the type-B NG mode is shown in Fig.~\ref{fig:DR}\,(b).
Taking account of one diffusive type-A NG mode, we see that the total number of the gapless NG modes is smaller than that of broken symmetry, which is peculiar to type-B NG modes.

\sect{Summary and discussion}
We have developed the EFT for the NG modes associated with SSB taking place in nonequilibrium open systems based on the coset construction.
The derived effective Lagrangian enables us to describe the semi-classical dynamics of the diffusive NG modes, which can be applied to general open systems from open quantum system including non-Hermitian quantum systems to classical stochastic systems.
As an application, we discussed the diffusive dispersion relation for type-A and type-B NG modes in two possible examples in condensed matter and high-energy physics.

Let us comment on some future problems.
While we only investigate the dispersion relation of the diffusive NG modes with vanishing $\zeta_{t/s}$-terms, the constructed effective Lagrangian \eqref{eq:Leff1} contains much more information on many-body processes of the diffusive NG modes.
It is interesting to investigate such information, e.g., low-energy theorems and possible instability caused by $\zeta_{t/s}$-terms based on our formalism.
It would also be important to understand how the loop correction from NG fields affects the low-energy behavior of systems like the Mermin-Wagner theorem~\cite{Mermin:1966fe,Hohenberg:1967zz,Coleman:1973ci}.
Also, it is worthwhile clarifying an additional constraint, known as dynamical KMS/thermal symmetry~\cite{Sieberer:2015hba,Sieberer:2015svu,Crossley:2015evo,Glorioso:2017fpd,Aron:2017spi} in thermal systems.
Another interesting direction is to consider the SSB of spacetime symmetry in open systems.
While the driven-dissipative BEC discussed in this Letter provides such a simplest example of time crystal, more general models may cause an instability towards the pattern formation~\cite{Kuramoto,MoriKuramoto}.
We left these problems and wide applications of the developed EFT in many physical systems from cold atomic, condensed matter, high-energy, and active matter systems, as future works.

\acknowledgments

The authors thank Yoshimasa Hidaka for critically useful discussions and 
invaluable comments.
M.H thanks Keisuke Fujii, Tetsuo Hatsuda and Yuki Minami for useful discussions.
M.H was supported by Japan Society of Promotion of Science (JSPS) 
Grant-in-Aid for Scientific Research (KAKENHI) Grant Numbers 18H01217, 
and the Special Postdoctoral Researchers Program at RIKEN. 
S.K. is supported in part by the Senshu Scholarship Foundation. 
T.N. is in part supported by JSPS KAKENHI Grant Numbers JP17H02894 and JP18K13539, and MEXT KAKENHI Grant Number JP18H04352. 
A.O. is supported by JSPS Overseas Research Fellowships. 
This work is partially supported by the Ministry of Education, Culture, Sports, Science, and Technology(MEXT)-Supported Program for the Strategic Research Foundation at Private Universities ``Topological Science'' (Grant No. S1511006), and the RIKEN iTHEMS Program, in particular iTHEMS STAMP working group.

\bibliography{coset_open}

\appendix
\pagebreak
\widetext
\begin{center}
\textbf{\large Supplemental Materials: 
\vspace{5pt} \\
Effective Lagrangian for Nambu-Goldstone modes in nonequilibrium open systems
} 
\end{center}

\setcounter{equation}{0}
\setcounter{figure}{0}
\setcounter{table}{0}
\setcounter{page}{1}
\makeatletter
\renewcommand{\theequation}{S\arabic{equation}}
\renewcommand{\thefigure}{S\arabic{figure}}
\renewcommand{\bibnumfmt}[1]{[S#1]}

\section{Derivation of $G_R$-transformation rules \eqref{eq:GRTransfpi}}
\label{sec:TransformationRule}
We here give the derivation of $G_R$-transformation rules of the NG fields given in Eq.~\eqref{eq:GRTransfpi}.
For that purpose, let us explicitly write the $G_A$-transformation given in Eq.~\eqref{eq:G12Transf} as
\begin{align}
 \label{G_A_pre}
 \xi_{1,2} (\bpi)
 \to
 \xi_{1,2} (\bpi')
 = e^{\pm \im \theta} e^{\im \pi_R} e^{\pm \im \pi_A/2}e^{- \im \beta_{1,2} (\bpi,\theta)},
\end{align}
where $\theta = \theta^I X_I$ denotes the transformation parameter with the pullback $\beta (\bpi,\theta) = \beta^\alpha (\bpi,\theta) X_\alpha$.
To determine the pullback $\beta (\bpi,\theta)$, based on the power counting scheme in which $\pi_A,~\theta$ and $\beta_{1,2}$ are all counted as $O (\hbar)$, we expand $\xi_{1,2}(\bpi)$ and their transformations \eqref{G_A_pre} as follows:
\begin{equation}
 \begin{split}
 \xi_{1,2} (\bpi)
 &
 =e^{\im\pi_R} \pm \frac{\im}{2} e^{\im\pi_R} \pi_A + O(\hbar^2)
 \\
 \xi_{1,2}(\bpi')
 &
 =
\left[
 e^{\im \pi_R}
 + O (\hbar^2)\right]
 \pm \frac{\im}{2} e^{\im \pi_R}
 \left[
  \pi_A + 2 e^{-\im \pi_R} \theta \, e^{\im \pi_R} 
  \mp 2 \beta_{1,2} + O (\hbar^3)
  \right]
  .
 \end{split}
\end{equation}
Comparing these, we can express the transformation rules for $\pi_R$ and $\pi_A$ as
\begin{equation}
 \begin{cases}
  \pi'_R = \pi_R + O (\hbar^2) ,
  \\
  \pi_A' = \pi_A + 2 e^{-\im \pi_R} \theta e^{\im \pi_R} 
  - 2 \beta_{1} + O (\hbar^3)
  = \pi_A + 2 e^{-\im \pi_R} \theta e^{\im \pi_R} 
  + 2 \beta_{2} + O (\hbar^3) .
 \end{cases}
\end{equation}
We therefore identify that the pullback $\beta_{1,2}$ takes the following form in our parametrization:
\begin{align}
 \beta_{1} 
 = - \beta_2
 = \left[e^{-\im\pi_R} \theta e^{\im\pi_R}\right]_{\para},
\end{align}
where $A_\para \equiv \tr (X_\alpha A) X^\alpha$ is the unbroken component of $A$.
This completes the semi-classical transformation rule of the NG fields given in Eq.~\eqref{eq:GRTransfpi}.

\section{Derivation of the matching condition \eqref{eq:matching1} and \eqref{eq:matching2}}
Since the first two matching conditions in Eq.~\eqref{eq:matching1} immediately follows from the transformation rule \eqref{eq:Transf}, we here provide the derivation of the others in Eqs.~\eqref{eq:matching1}-\eqref{eq:matching2}.
By using the retarded Green's function for NG modes given in Eq.~\eqref{eq:prop:RA}, we can directly evaluate $G_{J_{a,R}^0 J_{b,A}^0} (\omega,\bk)$ as follows:
\begin{equation}
 \begin{split}
  G_{J_{a,R}^0 J_{b,A}^0} (\omega,\bk)
  &= \int \diff^d x e^{\im \omega (t-t') - \im \bk \cdot (\bx-\bx')}
  \baverage{
  F \big( 
  - \im \omega f^{-2}g_{ac} - m^{\alpha'} f_{\alpha' ac} \big) \pi_R^c (x) 
  F \big( \im \omega g_{db}^t - \rho_{db}  \big) \pi_A^d (x')
  }
  \\
  &= F^2 \big( 
  - \im \omega f^{-2}g_{ac} - m^{\alpha'} f_{\alpha' ac} \big) 
   \big( \im \omega g_{db}^t - \rho_{db}  \big)
  \frac{-\im}{- g_{cd}^t \omega^2 - \im \rho_{cd} \omega + g_{cd}^s \bk^2} 
   \\
  &\xrightarrow{\bk \to 0}
  F^2 \big( 
  \im \omega f^{-2}g_{ac} + m^{\alpha'} f_{\alpha' ac} \big) 
  \big( \im \omega g_{db}^t - \rho_{db}  \big)
  \frac{1}{\im \omega g_{cd}^t - \rho_{cd} } 
  \frac{1}{\omega}
  \\
  &=
  F^2 \big( 
  \im f^{-2}g_{ab} + \omega^{-1} m^{\alpha'} f_{\alpha' ab} \big) ,
 \end{split}
\end{equation}
which gives the matching condition for $f$, or the third equation in the matching condition \eqref{eq:matching1}. 

We can show the remaining ones in Eq.~\eqref{eq:matching2} in the similar manner if we notice that the equation of motion for $\pi_R^a$ together with the definition of $J_{a,R}^\mu$ in Eq.~\eqref{eq:Currents}, brings about
\begin{equation}
 \partial_\mu J_{a,R}^\mu
  = F
  \left( 
   \zeta_t^{\alpha'} f_{\alpha' ab} \partial_0^2 
   - \gamma g_{ab} \partial_0 
   + \zeta_s^{\alpha'} f_{\alpha' ab} \bnab^2
  \right) \pi_R^b
  + \im F g_{ab} \pi_A^b.
\end{equation}
This equation is a manifestation of the non-conserving feature of open systems.
Then, we can evaluate the low-frequency/wave number behavior of $C^\nu_{ab} (\omega,\bk)$ in the similar manner with 
$G_{J_{a,R}^0 J_{b,A}^0} (\omega,\bk)$, using the retarded Green's function of NG modes. 
Noting that $\average{\pi_A^a(x) \pi_A^b (x')} = 0$ thanks to the unitarity condition \eqref{eq:SK1}, we can evaluate the time-component $C^0_{ab} (\omega,\bk)$ as
\begin{equation}
 \begin{split}
  C^0_{ab} (\omega,\bk) 
  &\equiv 
    \int \diff^d x e^{\im \omega(t-t') - \im \bk \cdot (\bx - \bx')}
  \baverage{ F ( - \omega^2 \zeta_t^{\alpha'} f_{\alpha' ac} 
  + \im \omega \gamma g_{ac} 
  - \bk^2 \zeta_s^{\alpha'} f_{\alpha' ac} ) \pi_R^c (x) 
  F (\im \omega g_{db}^t - \rho_{db} ) \pi_A^d (x')}
  \\
  &= F^2 
  ( - \omega^2 \zeta_t^{\alpha'} f_{\alpha' ac} 
  + \im \omega \gamma g_{ac} 
  - \bk^2 \zeta_s^{\alpha'} f_{\alpha' ac} ) 
  (\im \omega g_{db}^t - \rho_{db} ) 
  \frac{-\im}{- g_{cd}^t \omega^2 - \im \rho_{cd} \omega + g_{cd}^s \bk^2} 
  \\
  &\xrightarrow{\bk \to 0}
  F^2 
  ( - \omega^2 \zeta_t^{\alpha'} f_{\alpha' ac} 
  + \im \omega \gamma g_{ac} ) 
  (\im \omega g_{db}^t - \rho_{db} ) 
  \frac{1}{\im \omega g_{cd}^t - \rho_{cd} } \frac{-\im}{\im \omega}
  \\
  &= F^2 
  ( \omega \zeta_t^{\alpha'} f_{\alpha' ab} - \im \gamma g_{ab} ) ,
 \end{split}
\end{equation}
whose $\omega \to 0$ limit and the derivative with respect to $\omega$ give the matching conditions for $\gamma$ and $\zeta_t$---the first two equations in Eq.~\eqref{eq:matching2}.
Also, the spatial component $C^i_{ab} (\omega,\bk)$ is similarly evaluated as
\begin{equation}
 \begin{split}
  C^i_{ab} (\omega,\bk) 
  &\equiv 
  \int \diff^d x e^{\im \omega(t-t') - \im \bk \cdot (\bx - \bx')}
  \baverage{F ( - \omega^2 \zeta_t^{\alpha'} f_{\alpha' ac} 
  + \im \omega \gamma g_{ac} 
  - \bk^2 \zeta_s^{\alpha'} f_{\alpha' ac} ) \pi_R^c (x) 
  F (+\im k^i) g_{db}^s \pi_A^d (x')}
  \\
  &= \im F^2 
  ( - \omega^2 \zeta_t^{\alpha'} f_{\alpha' ac} 
  + \im \omega \gamma g_{ac} 
  - \bk^2 \zeta_s^{\alpha'} f_{\alpha' ac} ) k^i g_{db}^s 
  \frac{-\im}{- g_{cd}^t \omega^2 - \im \rho_{cd} \omega + g_{cd}^s \bk^2} 
  \\
  &\xrightarrow{\omega \to 0}
  - F^2 \zeta_s^{\alpha'} f_{\alpha' ab} k^i 
  ,
 \end{split}
\end{equation}
which gives the matching condition for $\zeta_s$---the third equation in Eq.~\eqref{eq:matching2}.

\section{Diffusive type-A NG mode in driven-dissipative Bose-Einstein condensate}
We here provide a detailed analysis of the type-A NG mode followed by the presence of the driven-dissipative Bose-Einstein condensate.
Our starting point is the Schwinger-Keldysh Lagrangian 
\begin{equation}
 \begin{split}
  \Lcal_{\open}
  &= 
  \bigg[\im \phi_A^\dag \partial_0 \phi_R
  - \frac{1}{2m} \bnab \phi_A^\dag \bnab \phi_R 
  +\phi_A^\dag 
  \Big( \mu + \im \kappa  - (g - \im \gamma_0) |\phi_R|^2 \Big) 
  \phi_R\bigg]+(\mathrm{h .c.})
  + \frac{\im (A + 4 \gamma_0 |\phi_R|^2)}{2} |\phi_A|^2 ,
  \label{eq:DDBEC}
 \end{split}
\end{equation}
where we defined the Keldysh-basis fields $\phi_R = (\phi_1 + \phi_2)/2$ and $\phi_A \equiv \phi_1 - \phi_2$.
We truncated the action at the quadratic order with respect to $A$-type field, which is regarded as the semiclassical approximation to open quantum systems described by the Lindblad equation (See, e.\,g., Ref.~\cite{Sieberer:2015svu} for a review).
Here $(\mathrm{h .c.})$ represents the Hermitian conjugate, and $\kappa < 0$ and $\gamma_0 >0$ denote a driven particle-injection term
and a dissipative term describing a nonlinear particle loss.

Using infinitesimal parameters $\epsilon$ for time-translational symmetry $\Rbb_t$, and $\theta$ for internal $U(1)$ symmetry, we define infinitesimal $G_A$ and $G_R$ transformations in the semiclassical regime as
\begin{equation}
 \begin{cases}
  \delta_A \phi_R = 
  \epsilon_A \partial_0 \phi_R + \im \theta_A \phi_R,
  \\
  \delta_A \phi_A = 
  \epsilon_A \partial_0 \phi_A + \im \theta_A \phi_A,
  \\
 \end{cases}
 \mathrm{and} \quad
  \begin{cases}
   \delta_R \phi_R = 
   0,
   \vspace{3pt} \\
   \delta_R \phi_A = 
   \epsilon_R \partial_0 \phi_R + \im \theta_R \phi_R,
   \\
  \end{cases}
\end{equation}
where only $G_A$-symmetry is respected in Eq.~\eqref{eq:DDBEC}. 

Let us then investigate a homogeneous symmetry breaking solution $\ovl{\phi}_R$ and $\ovl{\phi}_A$, on the top of which the type-A diffusive NG mode appears.
Noting $\ovl{\phi}_A = 0$ resulting from the unitarity condition, we obtain the mean-field equation of motion for $\ovl{\phi}_R$:
\begin{equation}
 \Big( \im \partial_0  + \mu + \im \kappa 
  - (g - \im \gamma_0) |\ovl{\phi}_R|^2 \Big) \ovl{\phi}_R
  = 0.
\end{equation}
The driven particle injection $\kappa < 0$ make the trivial solution $\ovl{\phi}_R= 0 $ unstable, and as a result, there appears a nontrivial solution with a time-oscillating homogeneous condensate given by
\begin{equation}
 \ovl{\phi}_R = v e^{- \im \omega_0 t} 
  \with 
  v^2 = - \frac{\kappa}{\gamma_0}, \quad 
  \omega_0 = gv^2 - \mu.
  \label{eq:DDCondensate}
\end{equation}
Due to the time-oscillating condensate, this solution breaks $G_A = \mathbb{R}_t \times U(1)_M$ symmetry down to $H_A = U(1)_{M+t}$. 
Here $U(1)_{M+t}$ denotes the combination of the time-translation and $U(1)$ transformation satisfying $\omega_0 \epsilon_A - \theta_A = 0$, which let the driven-dissipative condensate \eqref{eq:DDCondensate} invariant.

Next, we consider the fluctuation on the top of the above symmetry breaking solution and derive the effective Lagrangian for the diffusive NG mode. 
The embedding of Nambu-Goldstone mode reads
\begin{equation}
 \begin{split}
  \phi_R 
  \simeq e^{\im\pi_R - \im \omega_0 t}
  \left[
  v+\sigma_R
  \right], 
  \quad\text{and}\quad \phi_A 
  \simeq e^{\im \pi_R - \im \omega_0 t}
  \left[
  \im (v+\sigma_R) \pi_A + \sigma_A 
  \right] \,,
 \end{split}
 \label{embedding}
\end{equation}
where $\pi_{R,A}$ and $\sigma_{R,A}$ represent a phase (Nambu-Goldstone) and gapped amplitude field, respectively.
To see the semiclassical dynamics of those fields, we dropped the higher-order terms including more than two $A$-type fields.
Substituting the embedding \eqref{embedding} into Eq.~\eqref{eq:DDBEC} and focusing on the leading-order quadratic terms, we obtain the following result:
\begin{equation}
 \begin{split}
  \Lcal_{\open}
  &= - \frac{2v^2}{2m} \bnab \pi_A \bnab \pi_R 
  + \frac{\im \bar{A} v^2}{2} \pi_A^2
  \\
  &\quad
  - \frac{1}{2} 
  \begin{pmatrix}
   \sigma_R & \sigma_A 
  \end{pmatrix}
  \begin{pmatrix}
   0 & - \dfrac{1}{m} \bnab^2 + 4 gv^2 \\
   - \dfrac{1}{m} \bnab^2 + 4gv^2 & - \im \bar{A}
  \end{pmatrix}
  \begin{pmatrix}
   \sigma_R \\
   \sigma_A 
  \end{pmatrix}
  + 2v
  \begin{pmatrix}
   2 \gamma_0 v^2 \pi_A - \partial_0 \pi_A 
   & - \partial_0 \pi_R
  \end{pmatrix}
  \begin{pmatrix}
   \sigma_R \\
   \sigma_A
  \end{pmatrix}.
 \end{split}
\end{equation}
After integrating out the gapped mode $\sigma$ and taking the low-energy limit $\bnab^2/m \ll 4 g v^2$, we obtain the effective Lagrangian of the NG fields as
\begin{equation}
 \begin{split}
  \Lcal_{\eff}
  &= - \frac{2v^2}{2m} \bnab \pi_A \bnab \pi_R 
  + \frac{\im \bar{A} v^2}{2} \pi_A^2
  \\
  &\quad 
  - \frac{\im (2\im v)^2}{2} 
  \begin{pmatrix}
   2 \gamma_0 v^2 \pi_A - \partial_0 \pi_A 
   & -  \partial_0 \pi_R
  \end{pmatrix}
  \begin{pmatrix}
   0 & - \dfrac{\im}{m} \bnab^2 + 4 \im gv^2 \\
  - \dfrac{\im}{m} \bnab^2 + 4 \im gv^2 & \bar{A}
  \end{pmatrix}^{-1}
  \begin{pmatrix}
   2 \gamma_0 v^2 \pi_A - \partial_0 \pi_A  \\
   - \partial_0 \pi_R
  \end{pmatrix}
 \\
  &\simeq
  - \frac{2\gamma_0 v^2}{g} \pi_A \partial_0 \pi_R
  - \frac{2v^2}{2m} \bnab \pi_A \bnab \pi_R 
  + \frac{\im \bar{A} v^2}{2} \left( 1 + \frac{\gamma_0^2}{g^2} \right) \pi_A^2
 \\
 &= \frac{\im}{2}
 \begin{pmatrix}
  \pi_R & \pi_A
 \end{pmatrix}
 \begin{pmatrix}
  0 & \im \left( - \dfrac{\gamma_0 v^2}{g} \partial_0 
  - \dfrac{v^2}{m} \bnab^2 \right) 
  \\
  \im \left( \dfrac{\gamma_0 v^2}{g} \partial_0 
  - \dfrac{v^2}{m} \bnab^2 \right)  
  & \bar{A} v^2 \left( 1 + \dfrac{\gamma_0^2}{g^2} \right)
 \end{pmatrix}
 \begin{pmatrix}
  \pi_R \\
  \pi_A
 \end{pmatrix},
 \end{split}
\end{equation}
where we introduced $\bar{A} \equiv A+ 4 \gamma_0 v^2$.
The derived effective Lagrangian provides the following inverse retarded Green's function in the Fourier space:
\begin{equation}
 G_R^{-1} (\omega,\bk) = 
  - \im \frac{\gamma_0 v^2}{g} \omega + \frac{v^2}{m} \bk^2, 
\end{equation}
which results in the expected diffusive dispersion relation 
for the type-A NG mode
\begin{equation}
 \omega = - \im \frac{g}{m\gamma_0} \bk^2\,.
\end{equation}
We can also specify the $G_{A,R}$ transformation rules of NG modes as
\begin{equation}
 \begin{cases}
  \delta_A \pi_R = 
  \epsilon_A \partial_0 \pi_R -\omega_0 \epsilon_A+\theta_A, 
  \\
  \delta_A \pi_A = 
  \epsilon_A \partial_0 \pi_A,
  \\
 \end{cases}
 \mathrm{and} \quad
  \begin{cases}
   \delta_R \pi_R = 0,
   \vspace{3pt} \\
   \delta_R \pi_A = 
   \epsilon_R \partial_0 \pi_R -\omega_0\epsilon_R+\theta_R, 
  \end{cases}
\end{equation}
from which one see that $\pi_{R/A}$ nonlinearly transforms under the $G_{A/R}$ transformation while linearly transforms under the $G_{R/A}$ transformation.

\end{document}